\magnification=\magstep1

\def\oneandahalfspace{\baselineskip=\normalbaselineskip
  \multiply\baselineskip by 7 \divide\baselineskip by 4}

\catcode`\@=11
\def\vereq#1#2{\lower3pt\vbox{\baselineskip1.5pt \lineskip1.5pt
\ialign{$\m@th#1\hfill##\hfil$\crcr#2\crcr\sim\crcr}}}
\catcode`\@=12
\overfullrule=0pt

\oneandahalfspace 
 
\def\ref{\hangindent=50pt \hangafter=1 \noindent}
 
\magnification=\magstep1
\centerline{\bf Metallicity Dependence of the Cepheid Calibration}
\bigskip
\centerline{M. Sekiguchi$^{1}$ and M. Fukugita$^{2,1}$}
\vskip5mm
\centerline{$^1$ Institute for Cosmic Ray Research, University of Tokyo,
Tanashi, Tokyo 188, Japan}
\centerline{$^2$ Institute for Advanced Study, Princeton, NJ 08540, U. S. A.}
 
\vskip2cm
 
 
The metallicity dependence of the Cepheid period-luminosity (PL)
relation has been a controversial issue. 
The more commonly accepted view is that the PL relation 
is insensitive to metal abundance, while the period-luminosity-colour
relation is fairly sensitive, as indicated by stellar 
pulsation models$^{1-4}$ (see also refs. 5,6).
The direct observational evidence is scanty, however.
Freedman and Madore$^7$ have studied Cepheids in three
regions having different metallicity in M31 and concluded no 
statistically significant residual
differences in the distance moduli.
On the other hand, Gould$^8$ has claimed that the data of
Freedman and Madore are rather indicating a large metallicity dependence
$\Delta \mu/\Delta [{\rm Fe/H}]=0.88\pm0.16$, where $\mu$ is
the distance modulus derived from the Cepheid PL relation. 
A moderate metallicity
dependence $\Delta \mu/\Delta [{\rm Fe/H}]=0.44$ was also concluded 
in a recent report from EROS photometry
for the Magellanic Clouds$^9$. 

Recently an accurate measurement of metallicity has been carried out
for 23 Galactic Cepheids by Fry and Carney$^{10}$ (hereafter FC). This
work has given us an opportunity to examine the metallicity dependence of
the Cepheid PL relation and its calibrations. 
The most commonly used calibration of the
Galactic Cepheids is probably that of Feast and Walker$^{11}$ based on
$V$ band photometry and the distance based on the ZAMS fitting to the
clusters that contain the Cepheids.
Laney and Stobie$^{12}$ (hereafter LS) have slightly updated the Feast-Walker
distance and also provided a calibration of the Cepheid PL relation
for the near infrared $JHK$ colour bands$^{13}$ 
in addition to that for the
$V$ band (their result for $V$ band hardly differs from that of
Feast and Walker). The work of LS may perhaps be taken as the most 
accurate calibration of the Cepheid PL relation to date, hence giving the
most accurate distance to the LMC with the Cepheid method. 

The FC sample contains 12 Cepheids in common with the LS sample:
EV Sct, V340 Nor, WZ Sgr, V Cen, SW Vel, CV Mon, S Nor, U Sgr,
TW Nor, SV Vul, QZ Nor, and T Mon in the order of increasing
metallicity. We show in Fig. 1 an uncritical
display of the residual of the fit to the Cepheid PL relation as a function of
metallicity [Fe/H]. The residual is defined by 
$$\delta M_\lambda=M_\lambda-
(A_\lambda \log P + \phi_{\lambda PL}),~~~ \lambda=V, J, H, K\eqno{(1)}$$ 
with the coefficients 
given in LS: $(A, \phi)=(-2.874, -1.197)$ for $V$, $(-3.306,-1.971)$
for $J$, $(-3.421,-2.243)$ for $H$ and $(-3.443, -2.297)$ for $K$
(we take the first set of their parameters, but other 
choices hardly modify the plot of Fig. 1).  Virtually the same plot
is obtained with the data and the fit given in Feast and Walker for the
$V$ band. The error of the metallicity measurement of FC is $\pm(0.02-0.03)$
dex except for SV Vul ($\pm0.04$) and CV Mon ($\pm 0.06$).
This figure
shows a conspicuous correlation of $M_\lambda$ with metallicity.
The surprising fact is that this metallicity dependence is disturbingly
large $\delta M_{\lambda}/\delta [{\rm Fe/H}]=-(1.3-1.4)$, almost independent
of the colour band from $V$ to $K$. We note that one highly deviated 
point (TW Nor) suffers from a very large reddening correction
$E_{B-V}({\rm OB})=1.34$ and $M_V$ may not be accurately determined. The
leftmost point EV Sct and QZ Nor (the second point from the right)
are suspected to be overtone pulsation (FC).  If we remove these
three Cepheids (this means we remove all highly scattered points
in Fig. 1), the metallicity gradient is  
$\delta M_{\lambda}/\delta [{\rm Fe/H}]=-(1.7-2.1)$.
In order to underscore the colour independent nature of the offset
we plotted $\delta M_\lambda$ ($\lambda=V,J,H$) versus $\delta M_K$
in Fig. 2.

Now the question arises as to whether this large metallicity dependence 
represents that of the intrinsic Cepheid PL relation or it arises from some 
other artefacts of
procedures used in estimating $M_\lambda$.  Let us remember that the
absolute magnitude is calculated with
$$\mu_0=\mu_V-R({\rm OB})E_{B-V}({\rm OB}) \eqno{(2)}$$
and
$$M_\lambda=m_\lambda-(A_\lambda/A_V) R({\rm Cepheid})
  E({\rm Cepheid})-\mu_0 \eqno{(3)},$$
where the V band modulus $\mu_V$ is obtained by ZAMS fitting assuming
the cluster metallicity [Fe/H]=0 (see e.g., refs. 11, 14).
The ZAMS track with lower metallicity
generally yields fainter magnitude for given colour
(e.g., ref. 15)
and hence this is just opposite to
the trend observed in Fig. 1: a metallicity correction for 
the ZAMS fitting would steepen the slope. 
It is unlikely that the metallicity dependence arises from the 
absorption correction for Cepheids in eq. (3), because the
correlation seen in Fig. 1 is wavelength independent, whereas
reddening is not.  
Then we are left with two possibilities:

\noindent
(i) the Cepheid PL relation indeed has a large metallicity dependence, or

\noindent
(ii) the absorption correction for $\mu_0$ has a large [Fe/H] dependence.

Case (i), if it does not looks very likely, does not conflict 
with any other observations, when the
uncertainties in separating absorption corrections from the metallicity
effect and also of the metallicity measurements are considered. 
The indicated metallicity dependence, however, is by more than  
a factor 10 larger than
the prediction of the theoretical models of the Cepheid pulsation$^{1-4}$.
The colour-band independence of the offset at first glance also looks a 
little unusual
from a theoretical view point.  This is not impossible, however, 
if the {\it coefficient} of the pulsation equation 
($P$ as a function of mass and radius) 
is strongly affected by metal abundance,
in contrast to what stellar model calculations suggest.$^{1,3}$ 
We remark that the sign of the period shift is 
consistent with the Oosterhoff period shift for RR Lyr,$^{16,17}$ 
for which stellar models also predict
the metallicity dependence by a factor of $>10$ smaller than observed.$^{18}$
If case (i) is true, an
on-going Hubble Space Telescope project for studying the metallicity
dependence for Cepheids in M101$^{19}$ 
should detect 
the effect; the raw distance moduli derived from inner and
outer regions may differ by 1 mag after absorption correction. 

In order to examine the model prediction for the metallicity dependence
of the period-colour (PC) relation, we show in Fig. 3 the residual of
the fit of $(B-V)_0$ obtained by LS as a function of [Fe/H], and compare
it with a prediction of Stothers.$^3$  Unfortunately, the
scatter of data is too large that one can hardly conclude whether the
residual shows the metallicity dependence.  The only conclusion
one can draw from the figure is that the metallicity effect
on the PC relation, if any, is not too much larger than the
theoretical model predicts.  However, this does not preclude the
possibility that the pulsation equation is strongly affected by
metal abundance: the contribution of the metal-abundance dependent term
to the zero point of the PC relation is one order of magnitude 
smaller than that
to the zero point of the PL relation (see ref. 3). 
Namely, this test does not serve for us to choose among the two possibilities.

The colour independent nature may also be taken to be consistent
with case (ii).  Namely, either selective extinction $E_{B-V}({\rm OB})$,
which is usually estimated from cluster reddenings or space reddenings of 
stars close to Cepheid, or $R$ factor, which is set to be a constant, 
receives a large abundance effect,
although, this differs from our conventional belief that
the extinction correction for B stars does not depend so strongly on
metallicity, for line blanketing effect is smaller for high
temperature stars. Also somewhat strange with this case is that the extinction
correction in (1) does not tend to cancel against that in (2) for the $V$ band.
We have examined possible correlation between $E({\rm OB})$, $E({\rm Cepheid})$
or $\mu_V-\mu_0$ and [Fe/H], but have not found any apparent correlations 
between those quantities.

One may suspect a systematic error in FC in estimating metallicity.
One of the suspects may be in their estimate of temperature. 
We have checked that their values of [Fe/H] do not show any 
significant correlation with colour $(B-V)_0$,
indicating no obvious errors in the temperature estimate
that correlates with metallicity. Indirect, but more significant is 
Fig. 2 which demonstrates a strong correlation among
residuals of fitting in different colour bands, pointing towards a systematic 
problem in the Cepheid work itself, rather than in the estimate of 
metallicity of FC. (We remark that similar correlations in the residuals
are noted in refs. 8 and 9.)
Namely, possible systematic error of FC, if any, does not account for the
entire problem posed here.

Neither of the two cases discussed above looks very likely, but certainly 
not impossible.
If (i) is correct, the period-luminosity relation for the 
Cepheid is represented better with the addition of a metallicity
term, for example for the $V$ and $K$ bands, such as 
$$M_V=-1.256(\pm 0.046)-2.874\log P-2.15(\pm0.44)[{\rm Fe/H}]
     ~~~(\sigma=0.111) \eqno{(4a)}$$
$$M_K=-2.385(\pm 0.039)-3.443\log P-1.72(\pm0.38)[{\rm Fe/H}]
     ~~~(\sigma=0.095) \eqno{(4b)}$$
as obtained by fitting our
9 Cepheids (the slope is fixed to be that from LMC; EV Sct, QZ Nor  and 
TW Nor are excluded). These are compared with 
$$M_V=-1.144(\pm 0.077)-2.874\log P~~~~~(\sigma=0.232) \eqno{(5a)}$$
$$M_K=-2.295(\pm 0.063)-3.443\log P~~~~~(\sigma=0.189) \eqno{(5b)}$$
without the metallicity term (these equations agree with 
what are given in LS within the error). The equation with the metallicity
term yields the LMC distance
0.4 mag {\it smaller} than estimated in LS for average metal abundance
of the LMC, [Fe/H]=--0.3. If (ii) is correct the ``true'' PL relation 
after correcting for the metal abundance effect on $\mu_0$ is
given by the equation with the third term absorbed into eq. (4).
A comparison of (4) with (5) indicate that the LMC distance may be
$\approx$+0.1 mag 
{\it larger}.  This shows that one cannot determine
the distance accurately from the Cepheid PL relation unless the
metallicity effect is well controlled.  

A positive aspect, on the other hand, is that the intrinsic scatter
of the Cepheid PL relation should substantially be smaller than is believed
now once the metallicity correction is made.  After the metallicity correction
the dispersion of the PL relation becomes 0.08--0.11 mag depending on the
colour band, which is about a half the scatter obtained by LS. 

The present authors were unable to penetrate into details of the
numbers obtained by LS beyond the point which is explicit in their paper.  
It is, however, obvious that the best present data used for the Cepheid 
calibration, 
which is a cornerstone of the
extragalactic distance scale, contains an unusually large metallicity
dependence that has been ignored in the literature.  This would bring 
a significant uncertainty into the LMC distance; 
the error of the LMC distance from the Cepheid PL relation
remains uncertain as large as 
--0.4 to +0.1 mag, till this problem is solved.    

\vskip10mm
\noindent
{\bf Acknowledgements}
 
We would like to thank Bohdan Paczy\'nski, John Caldwell,
Herb Rood and Rob Kennicutt for valuable discussions and comments improving
the manuscript.
This work was partially supported by
Grant-in-Aid of the Ministry of
Education of Japan (09640336). 
M.F. wishes to acknowledge support from
the Fuji Xerox Corporation at Princeton.

\vfil\eject 
\noindent
{\bf References}
\medskip

\ref (1) I. Iben \& R. S. Tuggle, R. S.  ApJ, {\bf 173}, 135, 1970

\ref (2) I. Iben \& A. Renzini, Phys. Repts. {\bf 105}, 329, 1984

\ref (3) R. B. Stothers, ApJ {\bf 329}, 712, 1988

\ref (4) C. Chiosi, P. R. Wood \& N. Capitanio, ApJS, {\bf 86}, 541, 1993

\ref (5) S. C. B. Gascoigne, MNRAS, {\bf 166}, 25p, 1974

\ref (6) J. A. R. Caldwell \& I. M. Coulson, MNRAS, {\bf 218}, 223, 1986

\ref (7) W. L. Freedman \& B. F. Madore, ApJ, {\bf 365}, 186, 1990

\ref (8) A. Gould, ApJ, {\bf 426}, 542, 1994

\ref (9) D. D. Sasselov, et al., Astro-ph/9612216, submitted to A\&A, 
    1997

\ref (10) A. M. Fry, \& B. W. Carney, AJ, {\bf 113}, 1073, 1997

\ref (11) M. W. Feast \& A. R. Walker, ARA\&A, {\bf 25}, 345, 1987
 
\ref (12) C. D. Laney \& R. S. Stobie, MNRAS, {\bf 266}, 441, 1994

\ref (13) C. D. Laney \& R. S. Stobie, MNRAS, {\bf 263}, 921, 1993 
 
\ref (14) J. A. R. Caldwell, The Observatory, {\bf 103}, 244, 1983

\ref (15) D. A. VandenBerg \& T. J. Bridges, ApJ, {\bf 278}, 679, 1984

\ref (16) P. T. Oosterhoff, The Observatory, {\bf 62}, 104, 1939

\ref (17) H. Arp, AJ, {\bf 60}, 317, 1955

\ref (18) A. V. Sweigart, A. Renzini, \& A. Tornamb$\grave{\rm e}$ 
          ApJ, {\bf 312}, 762, 1987

\ref (19) D. D. Kelson, et al., ApJ, {\bf 463}, 26, 1996






 
 
 



 

 
 




 

\vfil\eject

\hskip-12mm
{\bf Figure captions}
\bigskip
 
\item{\rm Fig. 1} Residual of the fit to the Cepheid PL relation as
a function of metal abundance [Fe/H]. Here the residual is defined by
$\delta M_\lambda=M_\lambda-(A_\lambda \log P + \phi_{\lambda PL})$ 
($\lambda=V, J, H, K$) with the coefficients A and $\phi_{\lambda PL})$ 
taken from LS together with the data of $M_\lambda$.

\medskip
\item{\rm Fig. 2} Correlation of the residuals for different colour
bands.  $\delta M_V,\delta M_J$ and $\delta M_H$ are plotted against 
$\delta M_K$. 

\medskip
 
\item{\rm Fig. 3} Residual of the fit to the Cepheid period-colour ($B-V$) 
relation as a function of metal abundance. The residual is defined by
$\delta (B-V)_0=(B-V)_0-(C \log P + D)$. Both coefficients and data
are taken from LS.  The curve shows $\delta (B-V)_0 \simeq 8.4\delta Z$
as predicted by Stothers.$^3$ 

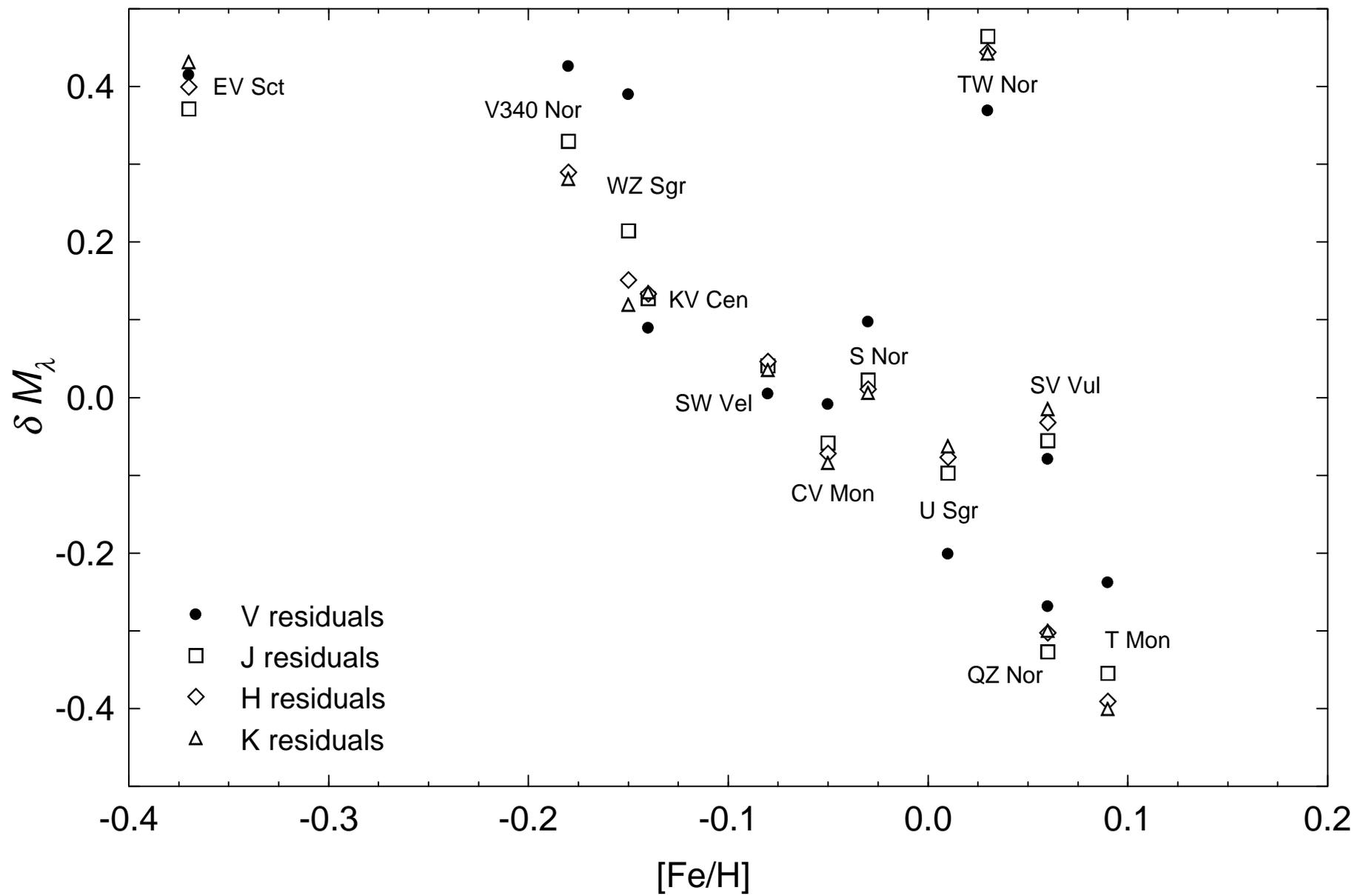
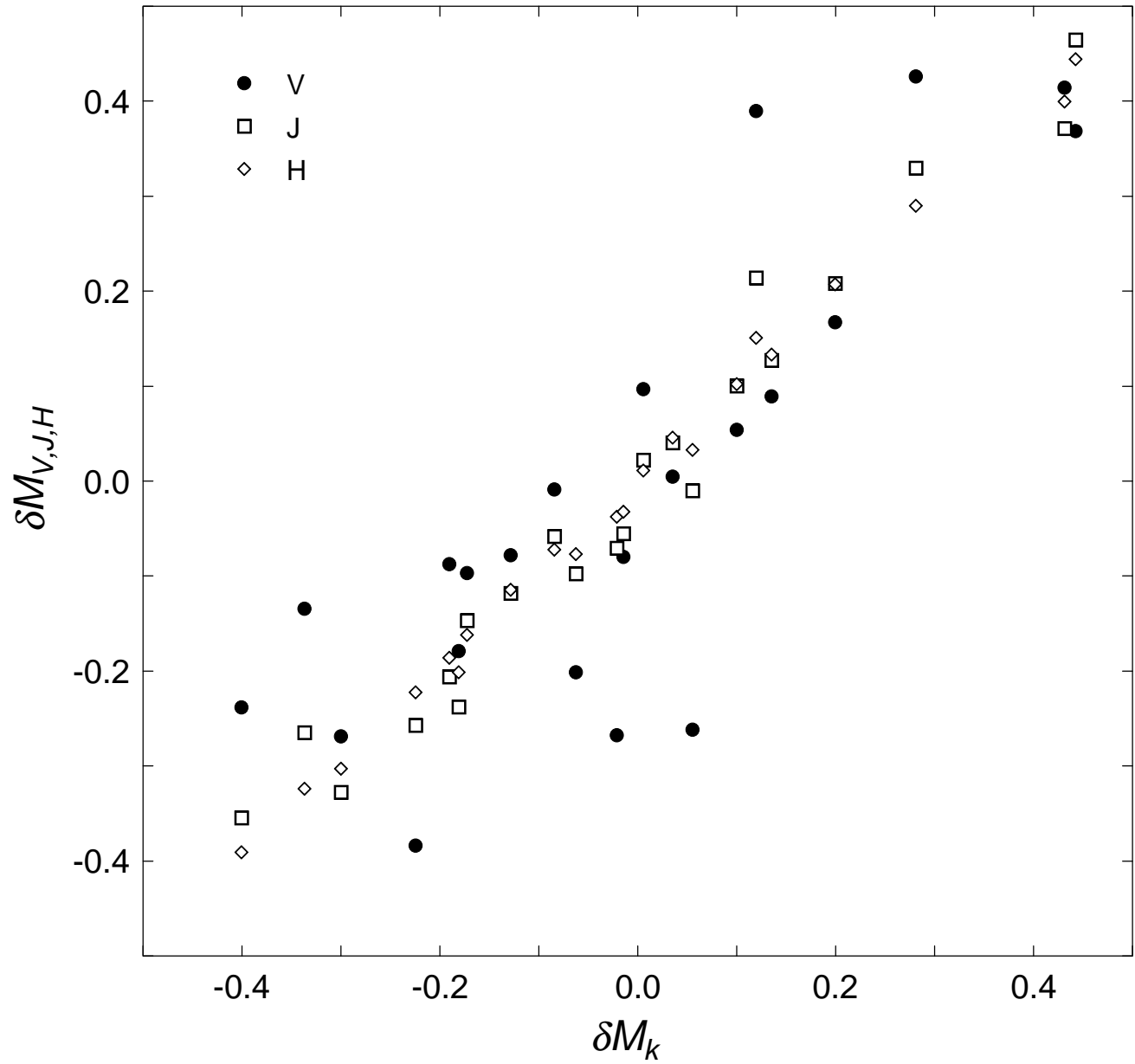
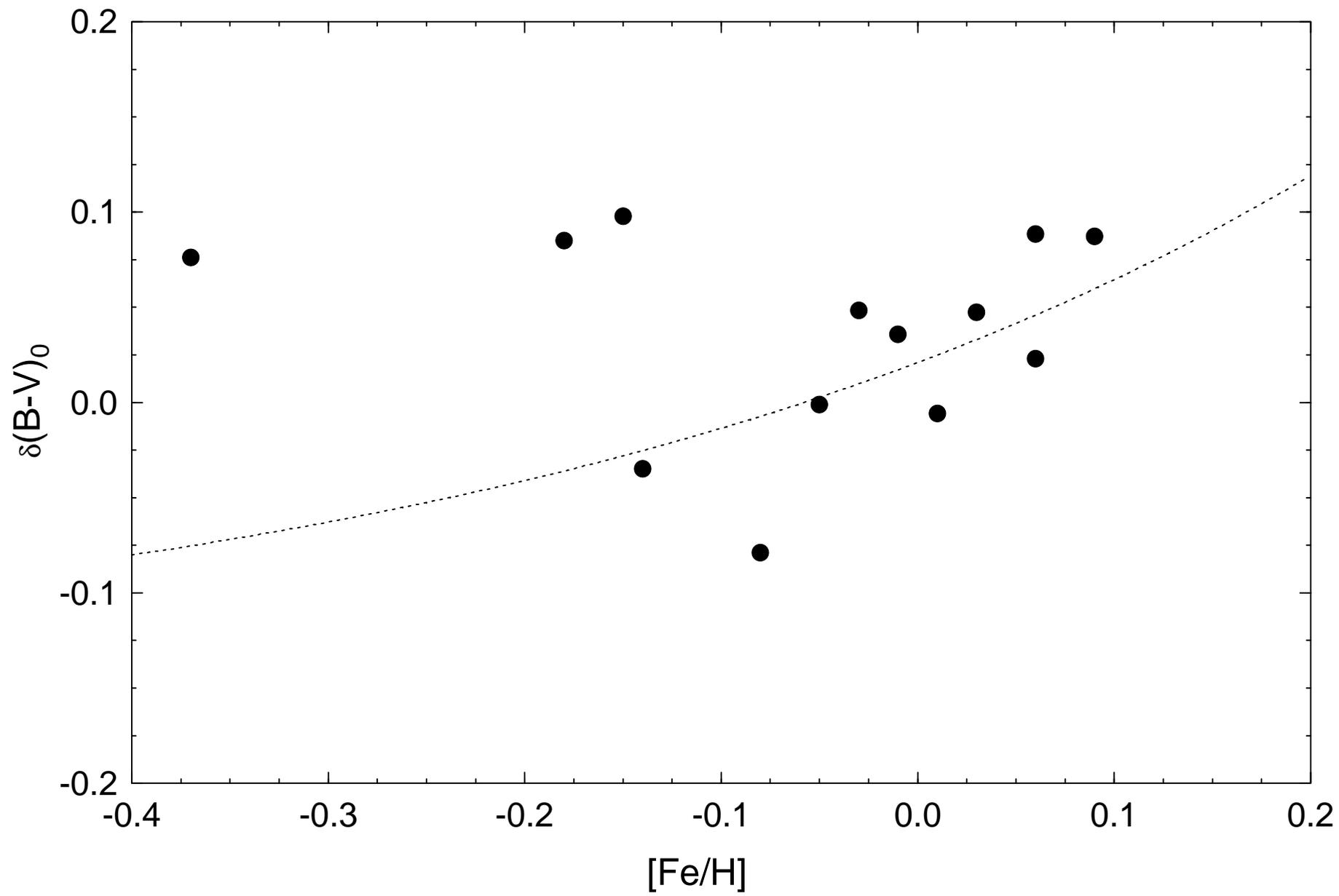
\bye